\documentclass[12pt]{iopart}
\usepackage{iopams}
\expandafter\let\csname equation*\endcsname\relax
\expandafter\let\csname endequation*\endcsname\relax
\usepackage{bm}
\usepackage[OT4]{fontenc}

\usepackage{graphicx}
\usepackage{xcolor}
\usepackage{epstopdf}

\newcommand{\kk}{\bm{k}}

\newcommand{\kp}{\bm{k}\!\cdot\!\bm{p}}

\newcommand{\mean}[1]{\langle #1 \rangle}

\begin{document}

\title[Phonon-assisted tunnelling of electrons in a QW-QD injection structure]{Phonon-assisted tunnelling of electrons in a quantum well-quantum dot injection structure}

\author{Adam Mielnik-Pyszczorski, Krzysztof Gawarecki and Pawe{\l} Machnikowski}

\address{Institute of Physics, Wroc{\l}aw University of
Technology, 50-370 Wroc{\l}aw, Poland}

\ead{Krzysztof.Gawarecki@pwr.wroc.pl} 

\begin{abstract}
We study theoretically phonon-assisted relaxation and tunnelling in a system composed of a quantum dot which is coupled to a quantum well. 
Within the $\bm{k} \cdot \bm{p}$ method combined with the L\"owdin elimination, we calculate the electron states. 
We calculate acoustic phonon-assisted relaxation rates between the states in the quantum well and in the quantum dot and study the resulting electron kinetics. We show that transition efficiency crucially depends on the system geometry.  We show also that under some conditions, transition efficiency can decrease with the temperature.
\end{abstract}


\maketitle

\section{Introduction}

Quantum dots (QDs) have been proposed for realization of various optical devices. In particular, QD lasers exhibit many advantages such as low threshold current \cite{liu99,asada86,shchekin02a,shchekin02b}, wide spectral tunability \cite{varangis00,nikitichev12}, or high temperature insensitivity \cite{varangis00,shchekin02b,shchekin02c,kirstaedter94,fathpour04,maksimov97}. However, a problem related with the concept of a QD laser is low carrier density inside the dot, which leads to low efficiency \cite{bhattacharya03}.
In order to avoid this problem, tunnel injection structures have been developed \cite{fathpour05}.
Due to high density of states, quantum wells (QWs) are good reservoirs providing carrier supplies for QDs. In a properly designed coupled QW-QD system, carriers can be injected with high speed \cite{bhattacharya03}, which considerably increases the optical efficiency. Carrier spectra as well as tunnel coupling have been widely investigated in double quantum dot systems \cite{bryant93,bayer01,korkusinski01,schliwa01,bester04,gawarecki10}. However, the energy structure in QW-QD system differs significantly from that case, due to the existence of the quasi-continuum of states in the QW. Recently, carrier states in such structures has been calculated within 8 band $\kp$ model on a 3D mesh under periodic boundary condition \cite{syperek12}.

The carrier kinetics in the QW-QD systems is also strongly affected by phonon-assisted processes which appear in a crystal environment. Carrier-phonon interaction leads to relaxation between states, which can involve carrier transfer (phonon-assisted tunnelling) between the two structures, that is, carrier capture to the QD. The essential role of phonons in the QW-QD injection process is confirmed by experiments \cite{rudno09a,rudno09b,syperek12,sek09b}, which indicate that the magnitude of the relaxation rate highly increases when the energy difference between the states of the QD and the QW becomes comparable with the energy of longitudinal optical (LO) phonons. Theoretically, carrier capture between structures of different dimensionality was studied for various systems and on different levels (Fermi golden rule \cite{ferreira1999pho}, Boltzmann kinetics \cite{nielsen2004man}, Green function formalism \cite{seebeck2005pol}, and full quantum kinetics
\cite{glanemann2005tra,reiter2006con,reiter2007spa}) involving LO phonons
\cite{ferreira1999pho,magnusdottir2002one,nielsen2004man} (also including two-phonon effects \cite{zhang2006rap,magnusdottir2002one}) and Coulomb dynamics
\cite{nielsen2004man}. The capture process involving tunnelling between a QW and a QD was analyzed within a model including LO phonons and Auger effects, based on a relatively simple model of wave functions \cite{chuang02,chang04}. On the other hand, it was shown \cite{markussen2006inf} that the exact shape of wave functions may be important for the correct calculation of the capture rates \cite{markussen2006inf}.

In this work, we study theoretically phonon-assisted tunnelling of electrons between a QW and a QD based on the realistic model of wave functions obtained by the $\bm{k} \cdot \bm{p}$ method for a strained structure. We take into account the electron coupling to acoustic phonons. We show that strain importantly changes the character of the lowest states in the QW (which cannot be accounted for in a simple model). Furthermore, we also investigate the carrier kinetics in the system. 

The paper is organized as follows. In Sec.~\ref{sec:model}, we present
the model. In Sec.~\ref{sec:results}, we discuss the results of the obtained electron states and carrier kinetics. Finally, concluding remarks and
discussion are contained in Sec.~\ref{sec:conclusion}. 

% We approximate the QW structure by a finite cylinder. Within an axial approximation we calculate the electron states using $\bm{k}\!\cdot\!\bm{p}$ method combined with the L\"owdin elimination \cite{lowdin51}. Then, phonon-assisted relaxation rates are derived within Fermi golden rule approach. Furthermore, using correlation expansion method, we find the electron kintetics.

\section{Model}
\label{sec:model}

\begin{figure}[tb]
\begin{center}
\includegraphics[width=50mm]{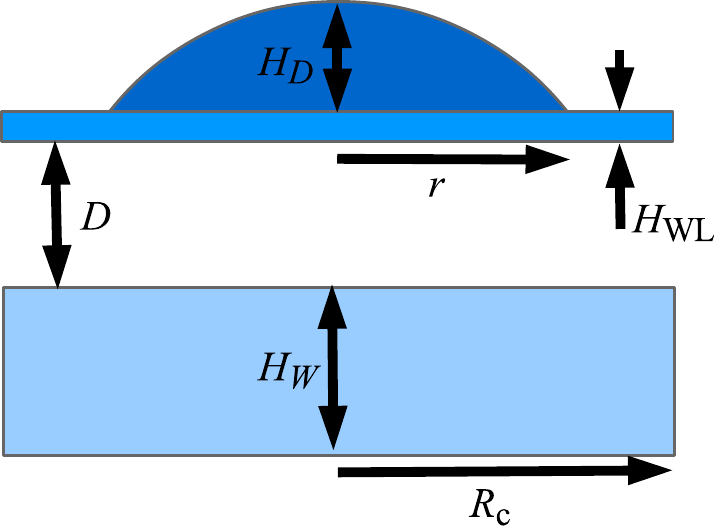}
\end{center}
\caption{\label{fig:schema}(Color online) The schematic cross-section of the system. }
\end{figure}

We investigate a vertically stacked system composed of a QW and a QD. The schematic picture of the system under consideration is shown in Fig.~\ref{fig:schema}.
We assume homogeneous alloying  $\mathrm{In_{0.62}\mathrm{Ga}_{0.38}As}$ inside the dot, $\mathrm{In_{0.2}\mathrm{Ga}_{0.8}As}$ in the QW and $\mathrm{In_{0.41}\mathrm{Ga}_{0.59}As}$ in the WL. The QW-layer thickness is set to $H_W=20 \mathrm{\ nm}$, the WL to $H_{\mathrm{WL}}=0.6 \mathrm{\ nm}$ and the QD high to $H_D=4.5 \mathrm{\ nm}$ (see Fig.~\ref{fig:schema} for the definition of the geometrical parameters). We assume also an axial symmetry of the system and perform the calculations in cylindrical coordinates $(\rho,\phi,z)$. Numerical computations are performed in a cylinder with the radius $Rc=300$~nm and height $Hc=80$~nm. The results have been verified for convergence with respect to the radius $R_c$.

The system is strained due to the lattice mismatch between InAs and GaAs. Following our previous work \cite{gawarecki10}, in order to calculate the elements of the strain tensor $\hat\epsilon$, we minimized the elastic energy of the system \cite{pryor98b,gawarecki10} in the continuous elasticity approach.
Because of the axial symmetry of the system, the wavefunctions can be represented in the form
\begin{equation}\label{ansatz}
\psi_{n}(\rho,z,\phi)=\frac{1}{\sqrt{2\pi}}
\varphi_{n}(\rho,z)e^{iM\phi},\nonumber
\end{equation}
where $M$ is the axial projection of the envelope angular momentum.
The local band structure is derived from the 8-band Hamiltonian with strain-induced terms (Bir-Pikus
Hamiltonian) using the L\"owdin elimination \cite{bahder90,lowdin51}. As a result, we obtain the effective Hamiltonian in the form 
\begin{eqnarray}
\label{schr}
H_{\mathrm{c}}  = & - \frac{1}{\rho} \frac{\partial}{\partial \rho} \rho
\frac{\hbar^{2}}{2m_{\bot}(\rho,z)} \frac{\partial}{\partial \rho}
-\frac{\partial}{\partial z}\frac{\hbar^{2}}{2m_{z}(\rho,z)} \frac{\partial}{\partial z} \nonumber\\
&+\frac{\hbar^{2} M^{2}}{2m_{\bot}(\rho,z) \rho^{2}} +E_{\mathrm{c}}(\rho,z),
\end{eqnarray}
with the conduction band edge 
\begin{displaymath}
E_{\mathrm{c}} (\rho,z)  =  E_{\mathrm{c}0}+a_{c} \mathrm{Tr} \{ \hat \epsilon \},
\end{displaymath}
where $ E_{\mathrm{c}0}$ is the unstrained bulk conduction band edge.
The in-plane component of the effective mass tensor takes the form 
\begin{displaymath}
m_{\bot}^{-1}(\rho,z)=m_{0}^{-1}\left( 
A'+\frac{E_{\mathrm{P}}}{2 E_{\mathrm{hh}}}
+\frac{E_{\mathrm{P}}}{6 E_{\mathrm{lh}}}
+\frac{E_{\mathrm{P}}}{3 E_{\mathrm{so}}} \right)
\end{displaymath}
and  the $z$ component of the effective mass is
\begin{displaymath}
m_{z}^{-1}(\rho,z)=m_{0}^{-1}\left( 
A'+\frac{2E_{\mathrm{P}}}{3 E_{\mathrm{lh}}}
+\frac{E_{\mathrm{P}}}{3 E_{\mathrm{so}}} \right),
\end{displaymath}
where \begin{displaymath}
A'= \frac{E_{p}(E_{g}+2\Delta/3)}{E_{g}(E_{g}+\Delta)}, 
\end{displaymath}
$E_{\mathrm{P}}$ is given by $2 m_{0} P^{2}/\hbar^{2}$ (where $P$ is a parameter proportional to the interband matrix transition element), $E_{\mathrm{g}}$ is the energy gap, and energy differences dependent on the position are defined as
\begin{eqnarray}
E_{\mathrm{hh}}  & =  E_{\mathrm{g}}+(a_{c}-a_{v}) \mathrm{Tr} \{ \hat \epsilon \}- b_{v} [\epsilon_{zz} - 0.5(\epsilon_{\rho \rho} +\epsilon_{\phi \phi} )],\nonumber \\
E_{\mathrm{lh}}  & =  E_{\mathrm{g}}+(a_{c}-a_{v}) \mathrm{Tr} \{ \hat \epsilon \} + b_{v} [\epsilon_{zz} - 0.5(\epsilon_{\rho \rho} +\epsilon_{\phi \phi} )],\nonumber\\
E_{\mathrm{so}}  & =  E_{\mathrm{g}}+(a_{c}-a_{v}) \mathrm{Tr} \{ \hat \epsilon \} +\Delta ,\nonumber
\end{eqnarray}
$a_{\mathrm{c}},a_{\mathrm{v}},b_{v}$ are the conduction and valence band deformation
potentials. The occupations of the electron states in the QW are given by the Fermi-Dirac distribution with the chemical potential $\mu$, which is related to the surface density of the electrons. For a given chemical potential we calculate the concentration of electrons $n_{e}$ as a sum over all the occupations in the QW divided by the cylinder base surface ($\pi R^{2}_{c}$).

The Hamiltonian of the system interacting with acoustic phonons is \cite{grodecka08a}
\begin{eqnarray}
H &= \sum_{n} \epsilon_{n} a^{\dagger}_{n} a_{n} + \sum_{\bm{k}\lambda} \hbar \omega_{\bm{k}\lambda} b^{\dagger}_{\bm{k}\lambda} b_{\bm{k}\lambda} \nonumber\\ &+\sum_{n,m,\bm{k}\lambda} F_{nm\lambda}(\kk)  (  b_{\bm{k}\lambda} + b^{\dagger}_{\bm{-k}\lambda})  a^{\dagger}_{n} a_{m}, \nonumber
\end{eqnarray}
where $\epsilon_{n}$ denotes the energy of the $n$-th state, $a_{n\lambda}^{\dag},a_{n\lambda}$ are the creation and annihilation operators for the electron $n$-th state respectively, $b_{\kk\lambda}^{\dag},b_{\kk\lambda}$ are operators of creation and annihilation of a phonon with the wave vector $\kk$ and phonon branch $\lambda$ and $F_{nm\lambda}(\kk)=F^{*}_{nm\lambda}(-\kk)$ is the electron-phonon coupling constant \cite{gawarecki10}. We find the kinetics of the electrons by solving the Heisenberg equation of motion,
\begin{displaymath}
\frac{d}{dt}  \langle a^{\dag}_{i} a_{i}  \rangle =  \frac{i}{\hbar} \langle [ H, a^{\dag}_{i} a_{i} ] \rangle,  
\end{displaymath}
where $\langle a^{\dag}_{i} a_{i}  \rangle \equiv f_{i}$ is the average occupation of the $i$-th state.

We perform calculations following the correlation expansion (CE) approach. The detailed derivation is given in the Appendix.  As a result, we obtain
\begin{eqnarray}\label{RE2}
\dot{f}_{i} &=  \sum_{j,\epsilon_{j}>\epsilon_{i}} \gamma_{ij} \left \{  f_{j} ( n_{B}(\omega_{ji}) + 1) -  f_{i} n_{B}(\omega_{ji}) - f_{i} f_{j} \right \} \nonumber \\
&+\sum_{j,\epsilon_{j}<\epsilon_{i}} \gamma_{ij}  \left \{  f_{j} n_{B}(\omega_{ij}) -  f_{i} (n_{B}(\omega_{ij}) + 1) - f_{i} f_{j}  \right \},
\end{eqnarray}
where $\omega_{ij}=(E_{i}-E_{j})/\hbar$ and $\gamma_{ij}$ is phonon-assisted relaxation rate given by \cite{gawarecki10} 
\begin{eqnarray}
\gamma_{ij}=2 \pi J_{ij}(\omega_{ji}),\nonumber
\end{eqnarray}
where $J_{ij}(\omega_{ji})$ is a phonon spectral density,
\begin{eqnarray}
J_{ij}(\omega_{ji})=\frac{1}{\hbar^2}\sum_{\mathbf{k}\lambda} \left|F_{ij\lambda}(\mathbf{k})\right|^2\left[ \delta(\omega_{ji}-\omega_{\mathbf{k}\lambda}) +\delta(\omega_{ji}+\omega_{\mathbf{k}\lambda}) \right],\nonumber
\end{eqnarray}
and $n_{B}$ is the Bose distribution. Following Ref.~\cite{gawarecki10}, we took into account  coupling to phonons by deformation potential (DP) as well as piezoelectric field (PE). In order to account how fast charge is flowing into the dot from the QW, we introduce the capture rate  as $\gamma_{0}=\sum_{i} \gamma_{0i} f_{i} ( n_{B}(\omega_{i0}) + 1)$ where we add all relaxation rates from the states in the QW to the ground state (localized in the QD). This procedure describes phonon-assisted relaxation properly if the state in the QD state is completely unoccupied. Otherwise, the Pauli blockade reduces the charge transfer. In consequence, in order to study the time evolution of the occupations, we numerically solve Eq.~(\ref{RE2}). 

The average number of electrons in the QD was found by $\langle N_{qd}\rangle = \sum_{i} f_{i} \eta_{i}$, where $\eta_{i}$ is the probability of finding electron in the $i$-th state in the upper half of the system. The details related with calculations are given in Appendix \ref{sec:appa}.

\section{Results}
\label{sec:results}

\begin{figure}[htb]
\begin{center}
\includegraphics[width=89mm]{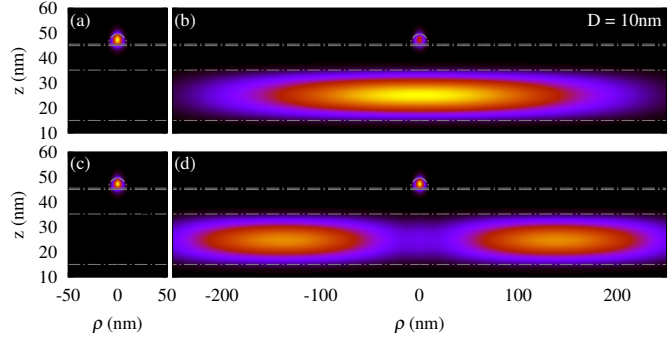}
\end{center}
\caption{\label{fig:dens}(Color online) The probability density for ground state and first excited state (the lowest state in the QW) in the presence of strain field (a,b) and in the case of neglected strain field (c,d).}
\end{figure}

We calculated single electron states in the considered structure. First, we investigate the influence of strain on electron states. We compared the probability density for the two lowest electron states in a hypothetical structure without strain (Fig.~\ref{fig:dens}(a,b)) and in a real strained structure (Fig.~\ref{fig:dens}(c,d)). In the former case, in order to have similar energy structure as the realistic one, we take a bulk effective mass and we adjust the conduction band edges to fixed values, constant within each structure (QD, QW, barrier). Ground states in both cases are localized in the QD and their character is the same. As shown in the Fig.~\ref{fig:dens}(b), if the strain field is disabled, the probability density of the lowest state in the QW (which has $M=0$) has a maximum at $\rho=0$. In the presence of strain (Fig.~\ref{fig:dens}(d)) the character of this state is different and the density forms a ring. This effect is caused by a repulsive potential generated by the strain field from the QD. However, for higher states in the QW (not shown here) this effect vanishes and for the sixth ($M=0$) state is no longer clearly visible. 

\begin{figure}[tb]
\begin{center}
\includegraphics[width=60mm]{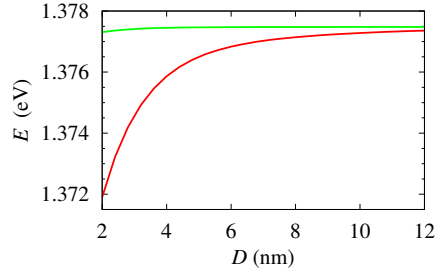}
\end{center}
\caption{\label{fig:energy}(Color online) The energies of two lowest states in the system as a function of the distance between the dot and the well. }
\end{figure}

\begin{figure}[tb]
\begin{center}
\includegraphics[width=89mm]{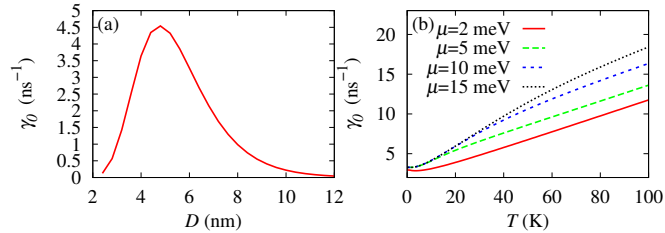}
\end{center}
\caption{\label{fig:ddep}(Color online) (a) The capture rate at $T=0$~K and $\mu=15$~meV ($n_{e}=1.7\cdot10^{11}\mathrm{cm}^{-2}$) as a function of the distance between the dot and the well. (b) Temperature dependence of the phonon-assisted relaxation rate at $D=4$~nm as a function of the temperature.}
\end{figure}

Fig.~\ref{fig:energy} presents the dependence of the two lowest electron energy levels on the distance $D$. The ground state is (mainly) localized in the dot and the first excited state is localized in the well. For a small distance, there is a strong tunnel coupling which leads to the large splitting between the energy of states in the QD and in the QW. 

% In Fig.~\ref{fig:energy}(b) the density of states in lower half of the system are shown. We present the states localized in the QW with $M=0$ (respectively: first, third and sixth). Near the dot symmetry axis, the dot generates also repulsive potential due to strain distribution. As a result, despite of $M=0$ character, the lowest QW states have no maximum at $\rho=0$. This effect vanishes for higher states and for the sixth state is not visible (as shown in Fig.~\ref{fig:energy}(b)).    

%
Next, we investigated the dependence of the capture rate $\gamma_{0}$ on the distance $D$. The results are shown in Fig.~\ref{fig:ddep}(a). 
The dependence is non-monotonic. On the one hand, for closely spaced structures the wavefunction overlap between the state localized in the QD and those from the QW is large, which is required for an efficient phonon-assisted relaxation process. On the other hand strong coupling leads to the large energy splitting, while the phonon spectral density at high frequencies is low  \cite{gawarecki12a,wu14}. Therefore, the efficiency of relaxation drops down. At large distances, the wave functions for the initial and final states have very small overlap, hence relaxation is also suppressed.

We studied also the temperature dependence of the capture rate.  In Fig.~\ref{fig:ddep}(b) the temperature dependences for several values of the chemical potential (corresponding to $n_{e}=2.2\cdot10^{10}$ to $n_{e}=1.7\cdot10^{11}\mathrm{cm}^{-2}$) are shown.  Observed dependence is linear at high temperatures because in this case, the leading term of the Bose distribution and Fermi-Dirac distribution is linear ($\sim kT$). The non-zero temperature on the one hand strongly increases the phonon spectral density but also reduces the occupations of electron states below the chemical potential (note that the lowest states in the QW give the predominant contribution in the relaxation process). The later effect is particularly important at small values of the chemical potential, as shown in Fig.~\ref{fig:ddep}(b): for $\mu=2$~meV ($n_{e}=2.2\cdot10^{10}$ $\mathrm{cm}^{-2}$) the relaxation rate decreases with the temperature (for small values of temperature).

\begin{figure}[tb]
\begin{center}
\includegraphics[width=89mm]{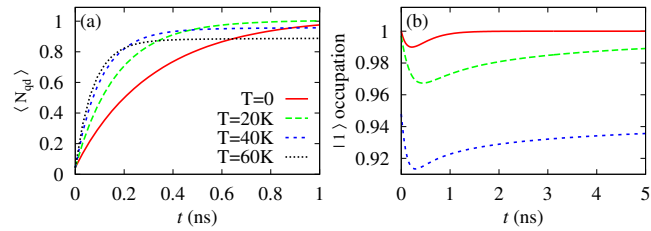}
\end{center}
\caption{\label{fig:kin}(Color online) (a) The time evolution of the average number of electrons in the QD. (b)  The time evolution of the lowest QW state occupation.}
\end{figure}

We solved Eq.~(\ref{RE2}) numerically and obtained the electron kinetics. In Fig.~\ref{fig:kin}(a) the time evolution of the average number of electrons in the QD, $\langle N_{qd}\rangle$ is shown. As the initial condition, we assume the zero occupation of the ground state (mainly localized in the dot) and Fermi-Dirac distribution in the well. Because of coupling between the dot and the well, the QW states are also partly localized in the dot and at $t=0$, the occupation of the QD is about $5$~\%. During the time evolution electrons from the well tunnel into the dot. The time dependence of $N_{qd}$ is nearly exponential with the initial slope similar to $\gamma_{0}$ (but slightly reduced due to the initial occupation). Increasing the temperature, on the one hand enhances this process, but also enhances the opposite effect (electrons can jump from the dot to the well). 
At high temperatures the occupation of the dot is reduced because of the thermal redistribution of occupations from the QD to the QW and also due to decreased occupation of the lowest QW states. 
Fig.~\ref{fig:kin}(b) shows the occupation of the lowest state in the QW. At $T=0$, the initial occupation is $1$, at the beginning of the evolution its value is decreasing because of phonon-assisted tunnelling to the dot. However, this effect is small and does not destroy the exponential character of $N_{qd}$ evolution.  However, higher states from the well also relax and full occupation is restored. In the case of non-zero temperature the initial occupation is lowered and electron can be excited to the higher state in the QW. In consequence the initial occupation is no longer recovered.    

\section{Conclusions}
\label{sec:conclusion}

We calculated the electron states in the system of the QD coupled with the QW. We shown that due to the strain-dependent repulsive potential, states in the QW are repelled from the dot axis.  We investigated the phonon-assisted tunnelling and relaxation. We obtained non-monotonic dependence of the relaxation rate on the distance between the QW and the QD. We studied the temperature dependence of the phonon transitions and we shown that value of the capture rate can decrease with temperature. Furthermore, we also investigated the electron kinetics. We obtained the exponential evolution of the average number of electrons in the dot. We also found a nonexponential evolution of the state occupations in the well.   

\ack

This work was supported in parts by the Foundation for Polish Science under the TEAM programme, co-financed by the European Regional Development Fund and by the grant No. 2012/05/N/ST3/03079 from the Polish National Sience Centre.

\appendix
\section{Calculation details} 
\label{sec:appa}

% \begin{figure}[tb]
% \begin{center}
% \includegraphics[width=50mm]{fig_conv}
% \end{center}
% \caption{\label{fig:conv}(Color online) Total value of the phonon-assisted relaxation rate at $D=4$~nm as a function of cylinder radius ($R_{c}$). }
% \end{figure}

In order to model the evolution of the electron occupations $f_{i}(t)$ we perform calculation within the CE approach. From the Heisenberg equation for $\mean{a^{\dag}_{i} a_{j}}$ we obtain
\begin{eqnarray}
 -i \hbar \frac{d}{dt}  \mean{a^{\dag}_{i} a_{j}} &=  ( \epsilon_{i} - \epsilon_{j} ) \mean{a^{\dag}_{i} a_{j}} \nonumber \\ 
  &+ \sum_{n,\bm{k}\lambda} F_{ni\lambda}(\kk) \left ( \mean{ a^{\dag}_{n} a_{j} b_{\kk\lambda}} + \mean{ a^{\dag}_{n} a_{j} b^{\dag}_{-\kk\lambda}} \right )   \nonumber \\ 
  &- \sum_{n,\bm{k}\lambda}  F^{*}_{nj\lambda}(\kk) \left( \mean{ a^{\dag}_{n} a_{i} b_{\kk\lambda}}^{*} + \mean{ a^{\dag}_{n} a_{i} b^{\dag}_{-\kk\lambda}}^{*} \right ). \nonumber
\end{eqnarray} 
In a similar way we calculate $\mean{ a^{\dag}_{n} a_{i} b_{\kk\lambda}}$. Here we assume fast relaxation of the reservoir which allows us to approximate $\mean{a^{\dag}_{i} a_{j} b^{\dag}_{\kk'\lambda} b_{\kk'\lambda}} \approx \mean{a^{\dag}_{i} a_{j}}
\mean{b^{\dag}_{\kk'\lambda}  b_{\kk'\lambda}}$. Furthermore, we neglected two-phonon processes $\mean{b^{\dag}_{\kk'\lambda}  b^{\dag}_{\kk'\lambda}} = \mean{b_{\kk'\lambda}  b_{\kk'\lambda}} = 0 $. After all these simplifications we obtain 
\begin{eqnarray}
-i \hbar \frac{d}{dt} \mean{ a^{\dag}_{n} a_{j} b_{\kk\lambda}} &=  \left(\epsilon_{n}-\epsilon_{j}-\hbar \omega_{\kk}\right) \mean{ a^{\dag}_{n} a_{j} b_{\kk\lambda}}   \nonumber \\ &+ \left .
\sum_{n'} F_{n'n\lambda}(-\kk) \mean{ a^{\dag}_{n'} a_{j}} n_{B}(\omega_{\kk}) \right . \nonumber \\ &- \left . 
\sum_{n'} F_{j n'\lambda}(-\kk) \mean{ a^{\dag}_{n} a_{n'}} (n_{B}(\omega_{\kk}) + 1)  \right . \nonumber \\ &- 
\sum_{n'm} F_{n'm\lambda}(-\kk) \mean{a^{\dag}_{n} a^{\dag}_{n'} a_{j} a_{m}},\nonumber
\end{eqnarray} 
where we took $\mean{b^{\dag}_{\kk'\lambda}  b_{\kk'\lambda}} = n_{B}(\omega_{\kk}) \delta_{\kk' \kk}$.
The electron correlations are accounted for within the Hartree-Fock approximation, $\langle a^{\dag}_{i} a^{\dag}_{n} a_{j} a_{m} \rangle \approx  \langle a^{\dag}_{i} a_{m} \rangle \langle a^{\dag}_{n} a_{j} \rangle - \langle a^{\dag}_{i} a_{j} \rangle \langle a^{\dag}_{n} a_{m} \rangle $.
%
%Then, we make the substitution $x_{ij}= \mean{a^{\dag}_{i} a_{j}} \exp{\left \{ %-(\epsilon_{i}-\epsilon_{j})/\hbar \right \} }$ and $y_{ij\kk}= \mean{a^{\dag}_{i} a_{j} %b_{\kk}} 
%\exp{ \left \{ -(\epsilon_{i}-\epsilon_{j}-\hbar \omega_{\kk})/\hbar \right \} }$.We obtain %closed set of equations for ${x}_{ij}$,${y}_{ij\kk}$ and $\dot{x}_{ij}$,$\dot{y}_{ij\kk}$. %After integration of $\dot{y}_{ij\kk}$ we take obtained $y_{ij\kk}$ into the equation for %$\dot{x}_{ij}$. After secular approximation we obtain
This yields a closed set of equations for $\mean{a^{\dag}_{i} a_{j}}$ and $\mean{a^{\dag}_{i} a_{j} b_{\kk}}$. The latter is then formally integrated and substituted to the former. Upon performing the Markov and secular approximations, we obtain
\begin{eqnarray}\label{RE}
\dot{f}_{i} &=  \sum_{j,\epsilon_{j}>\epsilon_{i}} \gamma_{ij} \left \{  f_{j} ( n_{B}(\omega_{ji}) + 1) -  f_{i} n_{B}(\omega_{ji}) - f_{i} f_{j} \right \} \nonumber \\
&+\sum_{j,\epsilon_{j}<\epsilon_{i}} \gamma_{ij}  \left \{  f_{j} n_{B}(\omega_{ji}) -  f_{i} (n_{B}(\omega_{ji}) + 1) - f_{i} f_{j}  \right \}.
\end{eqnarray}
This set of differential equations is solved numerically using the GSL library \cite{gsl}.

The Schr\"odinger equation with the Hamiltonian given by Eq.~(\ref{schr}) is solved numerically on a two-dimensional grid. The values of material parameters are taken from Ref.~\cite{gawarecki14a} except for $E_p$ which was taken $24.0$~eV for GaAs and $21.0$~eV for InAs. The eigenproblem has been solved using Lanczos method combined with the shift-invert spectral transformation where the linear set of equation is solved using the LIS library \cite{lis}. 

In order to check the validity of modeling an infinite well in a finite cylinder, we calculated the capture rate $\gamma_{0}$ as a function of cylinder radius $R_{c}$. We confirmed that $\gamma_{0}$ converge at $R_c=300$~nm. The reason for this convergence in spite of the quantized spectrum in the cylinder (as opposed to the actual continuum restored in the limit $R_c \rightarrow \infty$) is as follows: when the radius of the cylinder increases, the overlap between the wavefunction in the QD and those localized in the QW decreases like $\sim 1/R^{2}_{c}$. One the other hand, with increasing cylinder size, the density of QW states increases as $\sim R^{2}_{c}$. As a result, for a sufficiently large cylinder, convergence is reached.
\maketitle

\bibliographystyle{prsty}
\bibliography{abbr,quantum2}
\end{document}